\documentclass[letterpaper,twocolumn,english,aps,prl]{revtex4-1}
\usepackage{lmodern}
\usepackage{lmodern}
\usepackage[T1]{fontenc}
\usepackage[latin1]{inputenc}
\usepackage{color}
\usepackage{amsmath}
\usepackage{amssymb}
\usepackage{graphicx}

\makeatletter

\usepackage{babel}

\makeatother

\usepackage{babel}
\begin{document}

\title{Possible Chiral Topological Superconductivity in CrO$_{2}$ bilayers}

\author{Xu Dou, Kangjun Seo, and Bruno Uchoa$^{*}$ }

\affiliation{Department of Physics and Astronomy, University of Oklahoma, Norman,
OK 73069, USA}
\email{uchoa@ou.edu}

\selectlanguage{english}%

\date{\today}
\begin{abstract}
We address the possible emergence of spin triplet superconductivity
in CrO$_{2}$ bilayers, which are half-metals with fully spin-polarized
conducting bands. \textcolor{black}{Starting from a lattice model,
we show that chiral $p+ip$ states compete with non-chiral $p$-wave
ones.} At large doping, the $p+ip$ channel has a sequence of topological
phase transitions that can be tuned by gating effects and interaction
strength. Among several phases, we find chiral topological phases
having a single Majorana mode at the edge. \textcolor{black}{We show
that different topological superconducting phases could spontaneously
emerge in the vicinity of the van-Hove singularities of the band. }
\end{abstract}
\maketitle
\emph{Introduction.} Half-metals\emph{ }such as CrO$_{2}$ \cite{Yu,Soulen}
are promising materials for the prospect of emergent topological superconductivity.
By having a metallic Fermi surface with a single spin, they raise
the possibility of chiral superconductivity in the triplet channel
\cite{Pickett}, which is believed to occur only in a handful of systems
such as Sr$_{2}$RuO$_{4}$ \cite{Mackenzie}, which may have a spinfull
triplet state, UPt$_{3}$ and some heavy fermions superconductors
\cite{Kallin,Maeno}. A distintictive property of spin triplet chiral
topological superconductivity is the presence of Majorana fermions
propagating at the edges \cite{Read,Qi2,Qi,Lee,Alicea,Sato} and half-flux
quantum vortices \cite{Chung2,Jang} that can trap Majorana modes
\cite{Jackiw,Xu}. Majorana edge states were predicted to exist in
different heterostructures with strong spin-orbit coupling \cite{Fu,Akhmerov,fu2,Lutchyn,Chung,Li}
and may have been recently observed in an anomalous Hall insulator-superconductor
structure \cite{Qi3,He}.

In its most common form, CrO$_{2}$ is a three dimensional bulk material
with rutile structure \cite{Schwartz,Katsnelson}. It was recently
suggested \cite{Cai} that CrO$_{2}$/TiO$_{2}$ heterostructures
have fully spin polarized conduction bands over a wide energy window
around the Fermi level, and behave effectively as a two dimensional
(2D) crystal. In its simplest 2D form, CrO$_{2}$ will form a bilayer.
It is natural to ask if this material could spontaneously develop
2D chiral topological superconducting phases and host Majorana fermions
\cite{Fu}. 

We start from a lattice model for a single CrO$_{2}$ bilayer to address
the formation of spin triplet pairs\textcolor{red}{{} }\textcolor{black}{either
with $p$-wave or} $p_{x}+ip_{y}$ symmetry, which leads to a fully
gapped state. Due to the strong anisotropy of the gap, the superconducting
order has a line of quantum critical points as a function of both doping and 
coupling strength. 
In the
$p+ip$ state, we show that the system has an exotic sequence of topological
phase transitions, that could be tuned with gating effects. 
Different non-trivial topological phases may occur in the
vicinity of van-Hove singularities of the band, where the density
of states (DOS) diverges, allowing the possibility for both conventional
and purely electronic mechanisms. We suggest that
this system may provide an experimental realization of intrinsic 2D
chiral topological superconductivity in the triplet channel.

\begin{figure}[b]
\label{fig1} \includegraphics[width=0.92\linewidth]{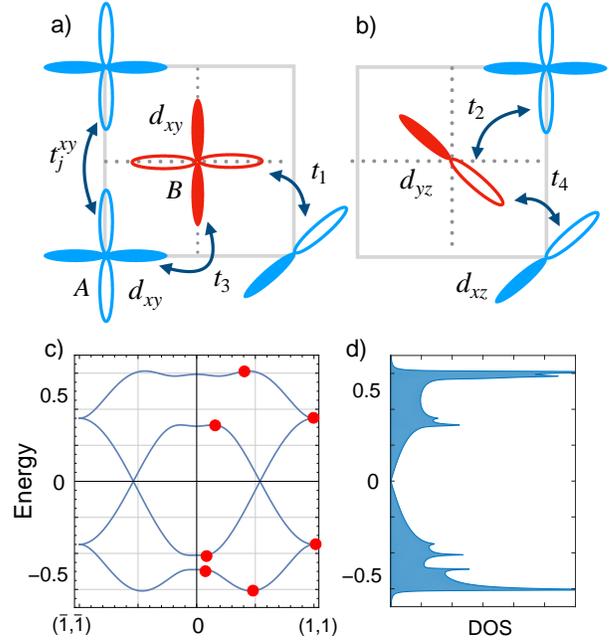} \caption{(Color online) top: Lattice of a CrO$_{2}$ bilayer, with $d_{xy}$
and $d_{xz}$ ($d_{yz}$) orbitals in sublattice $A$ (\textbf{$B$}).
The blue orbitals sit in the top layer ($A$ sites), and red orbitals
in the lower one ($B$ sites). Hopping energies are indicated by $t_{j}^{\alpha}$
for intra-orbital hopping between next-nearest sites, with $\alpha=xy,\,xz$
for $j=A$ and $\alpha=xy,\,yz$ for $j=B$, and $t_{i}$ ($i=1,2,3,4$)
for nearest neighbor hopping. c) Energy spectrum of the lattice model
along the diagonal $(1,1)$ direction. Energy axes in eV units. Red
dots indicate the location of van Hove singularities, where the DOS
(d) diverges logarithmically. }
\end{figure}

\emph{Lattice model.} In a bilayer system, the Cr atoms form two interpenetrating
square sublattices, $A$ and $B$, each one sitting on a different
layer. From above, the Cr atoms are arranged in a checkerboard pattern,
as shown in Fig. 1. Each site on sublattice $A$ ($B$) has two orbitals
with $d_{xy}$ and $d_{xz}$($d_{yz}$) symmetry. Nearest neighbors
(NN) hopping between a $d_{xy}$ orbital in sublattice $B$ with a
$d_{xz}$ orbital in sublattice $A$ has amplitude $t_{1}$ along
the the $(1,\bar{1})$ direction and zero along the $(1,1)$ direction
by symmetry. In the same way, NN hopping between a $d_{xy}$ orbital
in sublattice $A$ and with a $d_{yz}$ orbital in $B$ has amplitude
$t_{2}$ along the $(1,1)$ direction and zero along the other diagonal
in the $xy$ plane. Intra-orbital NN hopping is finite between $d_{xy}$
orbitals $(t_{3}$) but zero between $d_{xz}$ and $d_{yz}$ orbitals
($t_{4}$), which are othogonal to each other. Among next-nearest
neighbors (NNN), the dominant processes are described by intra-orbital
hoppings $t_{j}^{\alpha}$, with $\alpha=xy,\,xz$ for sites in sublattice
$j=A$ and $\alpha=xy,\,yz$ for $B$ sites.

The Hamiltonian can be described in a four component basis $\Psi=(\psi_{A,xy},\psi_{A,xz},\psi_{B,xy},\psi_{B,yz})$.
In momentum space, $\mathcal{H}_{0}=\sum_{\mathbf{q}}\Psi_{\mathbf{q}}^{\dagger}h(\mathbf{q})\Psi_{\mathbf{q}}$,
with \cite{Cai} 
\begin{align}
h(\mathbf{q})= & \left(\begin{array}{cc}
h_{A} & h_{AB}\\
h_{AB}^{\dagger} & h_{B}
\end{array}\right),\label{1}
\end{align}
where 
\begin{equation}
h_{A}=\left(\begin{array}{cc}
\epsilon_{A}^{xy}(\mathbf{q}) & 0\\
0 & \epsilon_{A}^{xz}(\mathbf{q})
\end{array}\right),\,h_{B}=\left(\begin{array}{cc}
\epsilon_{B}^{xy}(\mathbf{q}) & 0\\
0 & \epsilon_{B}^{yz}(\mathbf{q})
\end{array}\right).\label{eq:hAB}
\end{equation}
The diagonal terms incorporate NNN hopping processes, where $\epsilon_{j}^{\alpha}(\mathbf{q})=E_{j}^{\alpha}+4t_{j}^{\alpha}\mathrm{cos}q_{x}\mathrm{cos}q_{y}$,
with $E_{j}^{\alpha}$ a local potential on obital $\alpha$ in sublattice
$j$ and $q_{x,y}=\frac{1}{2}(k_{x}\mp k_{y})$ the momentum along
the two diagonal directions of the crystal. The off-diagonal terms
in (\ref{1}) describe the NN hopping terms illustrated in panels
a) and b) in Fig. 1, 
\begin{align}
h_{AB}= & \left(\begin{array}{cc}
-2t_{3}\sum_{\nu=x,y}\mathrm{cos}q_{\nu} & 2it_{1}\mathrm{sin}q_{y}\\
2it_{2}\mathrm{sin}q_{x} & -2t_{4}\sum_{\nu=x,y}\mathrm{cos}q_{\nu}
\end{array}\right),\label{eq:3}
\end{align}
where $t_{4}=0$ in the absence of spin-orbit coupling. 

The energy spectrum is shown in Fig. 1c, and has two sets of Dirac
points along the ($1,1$) and $(1,\bar{1})$ directions, respectively.
Enforcing the symmetries of the 2D lattice, namely roto-inversion
$S_{4}$ symmetry and mirror symmetry $M$ at the diagonal directions
of the unit cell, we adopt $t_{1}=-t_{2}\equiv t\sim0.3$eV as the
leading energy scale, and the set of parameters $t_{3}\sim t/30$,
$t_{j}^{xy}=-t_{A}^{xz}=-t_{B}^{yz}\sim t/3$, and $E_{j}^{xy}=-E_{A}^{xz}=-E_{B}^{yz}\sim t/6$.
We also \textcolor{black}{use $t_{4}\sim it/8$}, following ab initio
results \cite{note1,Cai}. The four band model breaks down near the
edge of the band, where states may hybridize with high energy bands.
We also assume that the bands are spinless. The resulting band structure
has several van Hove singularities at the saddle points, where the
density of states (DOS) diverges logarithmically, as depicted in Fig.
1d. In the vicinity of those points (red dots), the system can be
unstable towards superconductivity. 

\emph{Pairing Hamiltonian.} For spinless fermions, superconductivity
is allowed only in the triplet channel.\emph{ }The wavefunction of
the Cooper pairs is anti-symmetric under inversion, and hence only
states with odd angular momentum are allowed. \textcolor{black}{When
electrons pair accross the center of the Brillouin zone, the lowest
symmetry is in the $p$-wave channel, which can be induced by NN pairing.
$f$-wave pairing may be induced with NNN pairing only. This channel
is subdominant and will not be addressed. We consider the possible
instabilities of the lattice model both in the $p$-wave and in the
chiral $p+ip$ state, which can produce a full gap.} A conclusive
assessment of the stability of those states requires taking fluctuations
into account \cite{Furukawa,Nandkishore,Kiessel}, which will be considered
elsewhere.

For NN sites, the effective interaction term has the form 
\begin{equation}
\mathcal{H}_{\text{int}}=-\frac{1}{2}\sum_{\mathbf{r}\in\text{NN}}g^{\alpha\beta}\hat{n}_{i,\alpha}(\mathbf{r}_{i})\hat{n}_{j,\beta}(\mathbf{r}_{j})\label{eq:Hint}
\end{equation}
where $\hat{n}_{i,\alpha}=\psi_{i,\alpha}^{\dagger}\psi_{i,\alpha}$
is the density operator in orbital $\alpha$ on sublattice $i=A,\,B$,
$g^{\alpha\alpha}\equiv g_{1}>0$ is the intra-orbital coupling, and
$g^{xy,yz}=g^{xz,xy}\equiv g_{2}>0$ is the coupling in the inter-orbital
channel. The $p+ip$ pairing follows from the Ansatz on the lattice
$\Delta^{\alpha\beta}(\delta_{n})=g^{\alpha\beta}\langle\psi_{A,\alpha}(\mathbf{r})\psi_{B,\beta}(\mathbf{r}+\vec{\delta}_{n})\rangle\equiv\Delta^{\alpha\beta}\text{e}^{i\frac{\pi}{2}n},$
where $\vec{\delta}_{1,3}=\pm\frac{a}{2}(\hat{x}+\hat{y})$ and $\vec{\delta}_{2,4}=\pm\frac{a}{2}(\hat{x}-\hat{y})$
describe the four NN vectors, with $a$ the lattice constant. \textcolor{black}{For
$p$-wave pairing, we use the Ansatz} $\Delta^{\alpha\beta}(\pm\delta_{1,2})\equiv\pm\Delta^{\alpha\beta}$. 

Defining $\Delta^{\alpha\alpha}\equiv\Delta_{1}$ and $\Delta^{\alpha\beta}\equiv\Delta_{2}$
for intra-orbital and inter-orbital pairing respectively, the order
parameter in momentum space $\Delta_{i}^{C}(\mathbf{q})=\Delta_{i}^{C}(\sin q_{y}+i\sin q_{x})$
has chiral $p_{x}+ip_{y}$ symmetry, with $i=1,\,2$ and $q_{x,y}$
defined as above Eq. (\ref{eq:3}). \textcolor{black}{In the non-chiral
state,} $\Delta_{i}^{p}(\mathbf{q})=\Delta_{i}^{p}(\sin q_{y}+\sin q_{x})$
\textcolor{black}{has $p_{x}$ symmetry} \cite{note6}. At the mean
field level, Hamiltonian (\ref{1}) and (\ref{eq:Hint}) result in
the Bogoliubov-de Gennes (BdG) Hamiltonian $\mathcal{H}_{{\rm BdG}}=\sum_{\mathbf{k}\in BZ}\Phi_{\mathbf{q}}^{\dagger}h_{\text{{\rm BdG}}}(\mathbf{q})\Phi_{\mathbf{q}}$
with $\Phi_{\mathbf{q}}=(\Psi_{\mathbf{q}},\Psi_{-\mathbf{q}}^{\dagger})$,
which has the form 
\begin{align}
h_{{\rm BdG}}(\mathbf{q})= & \left(\begin{array}{cc}
h(\mathbf{q}) & \hat{\Delta}(\mathbf{q})\\
\hat{\Delta}^{\dagger}(\mathbf{q}) & -h^{T}(-\mathbf{q})
\end{array}\right),
\end{align}
where 
\begin{equation}
\hat{\Delta}(\mathbf{q})=\left(\begin{array}{cc}
0 & \Delta_{1}(\mathbf{q})\mathbf{1}+\Delta_{2}(\mathbf{q})\sigma_{x}\\
\Delta_{1}(\mathbf{q})\mathbf{1}+\Delta_{2}(\mathbf{q})\sigma_{x} & 0
\end{array}\right)\label{Delta2}
\end{equation}
is the pairing matrix, with $\sigma_{x}$ a Pauli matrix in the orbital
space. Minimization of the free energy $\mathcal{F}(\Delta_{1},\Delta_{2})=-T\text{tr}\sum_{\mathbf{k}}\ln e^{-h_{\text{BdG}}(\mathbf{k})/T}+\sum_{i=1,2}|\Delta_{i}|^{2}/g_{i}$
for a fixed chemical potential $\mu$ gives the zero temperature ($T=0$)
phase diagram shown in Fig. 2a, b as a function of the couplings $g_{1}$
and $g_{2}$. The two leading instabilities in the $p$-wave and chiral
$p+ip$ states compete with each other and are addressed below. 

\textcolor{black}{\emph{$p$-wave phase.}}\textcolor{black}{{} The vertical
line in fig. 2a describes a quantum phase transition at $g_{1}=\tilde{g}_{1c}(\mu)$
separating the normal phase (N) from the intra-orbital $p_{x}$ state
($p$SC I). At $g_{2}=\tilde{g}_{2,c}(\mu)$ (dashed line) the system
has a first order phase transition towards an inter-orbital $p$-wave
state ($p$SC II). The due to the anisotropy of the $p$-wave gap,
the curve $g=\tilde{g}_{1c}(\mu)$ depicted in Fig. 3a describes a
line quantum critical points with power law scaling \cite{note2}.
The intra-orbital channel $(\Delta_{1}^{p}\neq0$) dominates over
the inter-orbital one for all values of $\mu$. At the mean field
level, it is also the leading instability in the weak coupling regime
of the problem (which we define below), and subleading to the $p+ip$
state in the strong coupling sector, as indicated in Fig. 3a. }

\emph{$p+ip$ phase.} The inter-orbital channel $g_{2}$ may lead
to gapless chiral $p+ip$ superconductivity ($\Delta_{2}^{C}\neq0$)
shown in the red region, which is topologically trivial (Fig. 2a).
\textcolor{black}{This state dominates over the inter-orbital $p$SC
II phase, shown in Fig. 2b.} The dashed line around the gapless phase
in Fig. 2b describes a first order phase transition and sets the boundary
of the gapless $p+ip$ phase with the others at $g_{2}=g_{2c}(\mu)$.
The intra-orbital $p+ip$ pairing state ($\Delta_{1}^{C}\neq0$) on
the other hand is fully gapped and can be topological.

\begin{figure}[t]
\label{fig2} \includegraphics[width=0.97\linewidth]{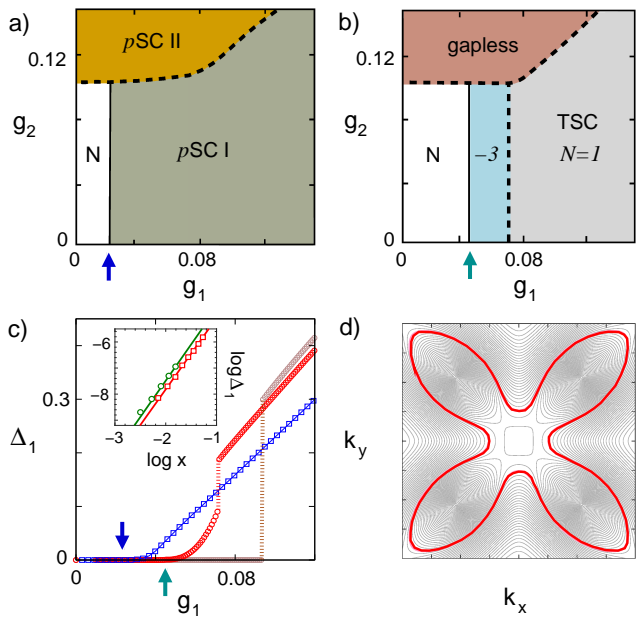}
\caption{(Color online) Phase diagram for a) $p$-wave and b) $p+ip$ state
for intra-orbital ($g_{1}$) and inter-orbital ($g_{2}$) couplings
at $\mu=4t/3=0.4$ eV. $g_{1}$ and $g_{2}$ in eV units. Green area:
intra-orbital $p$-wave ($p$SC I) for $g>\tilde{g}_{1c}(\mu)$. Yellow:
inter-orbital $p$-wave ($p$SC II). Gray area: strong coupling topological
phase (TSC) with Chern number $\mathcal{N}=1$; Light blue: gapped
weak coupling one ($\bar{g}_{1c}<g<g_{1c}$) with $\mathcal{N}=-3$
(see Fig. 3b). Light red: gapless $p+ip$, which is topologically
trivial. Dashed lines: first order phase transitions. Solid black
lines: second order transition (blue and green arrows). c) Scaling
of $\Delta_{1}$ vs $g_{1}$. Red and brown circles: $p+ip$ for $\mu=0.4$,
and $0$, respectively. Blue square: $p$-wave for $\mu=0.4$ eV.
Green and blue arrows: critical coupling $\bar{g}_{1c}\approx t/7=0.045$
eV and $\tilde{g}_{1c}\approx t/12=0.025$ eV respectively for $\mu=0.4$
eV. Inset: $\Delta_{1}^{C}$ vs $\text{log}x$, with $x=(1-\bar{g}_{1c}/g)$,
showing power law scaling in the chiral phase near $\bar{g}_{1c}(\mu)$.
Green dots: $\mu=0.3\,$eV. Red: $\mu=0.4$eV. d) Brillouin zone.
Red line: anisotropic Fermi surface at $\mu=0.33$ eV. }
\end{figure}

The gapped state has multiple minima that compete. The dashed vertical
line in Fig. 2b indicates a first order phase transition between the
weak and strong coupling topological phases (TSC) at $g_{1}=g_{1c}(\mu)$.
At this coupling, the superconducting order parameter $\Delta_{1}^{C}$
jumps (see Fig. 2c) and different gapped phases with distinct topological
numbers coexist. The resulting gap is very anisotropic around the
Fermi surface (Fig. 2d). In the weak coupling phase $\bar{g}_{1c}(\mu)<g<g_{1c}(\mu)$
shown in the light blue region in Fig. 2b, the intra-orbital chiral
gap $\Delta_{1}^{C}$ scales as a power law with the coupling for
fixed $\mu$, 
\begin{equation}
\Delta_{1}^{C}(g_{1})\propto(1-\bar{g}_{1c}/g)^{\beta},\label{eq:Delta2}
\end{equation}
with $\beta\approx2.7\pm0.1$ for $0.2\lesssim\mu\lesssim0.4$ eV
(see Fig. 2c inset). $\Delta_{1}^{C}$ vanishes at the critical coupling
$\bar{g}_{1c}$, where the system has a second order phase transition
to the normal state, indicated by the green arrows in Fig. 2. A qualitatively
similar behavior is also observed in the scaling of the intra-orbital
$p$-wave gap $\Delta_{1}^{p}$ near the critical coupling $\tilde{g}_{1}(\mu)$
(blue arrows in Fig 2) \cite{note2}. 

\begin{figure}[t]
\label{fig4-1} \includegraphics[width=1\columnwidth]{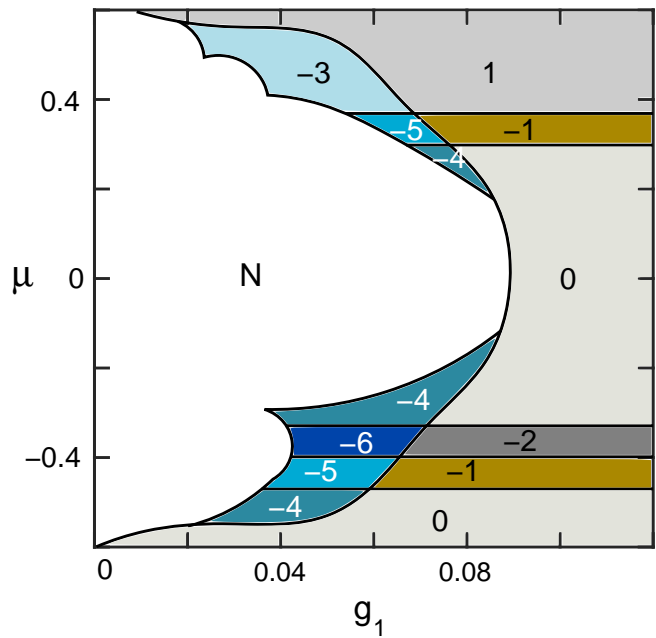}\caption{{\small{}{}(Color online) Mean field phase diagram as a function
of the chemical potential $\mu$ and intra-orbital pairing coupling
$g_{1}$, both in eV units. The phase diagram does not depend on $g_2$ for $g_2\lesssim 0.1$ eV (see Fig 2). 
 a) Normal phase (N), $p_{x}$ superconducting
phase (}\emph{\small{}p}{\small{}SC I), chiral $p+ip$ state (CSC)
and topological $p+ip$ phase (TSC). b) Possible topological phases
in the chiral $p+ip$ channel. The integers indicate the corresponding
BdG Chern number }\emph{\small{}$\mathcal{N}$}{\small{}. For fixed
$g_{1}$, the system has a sequence of topological phase transitions
near the van-Hove singularities of the band, where the topology of
the Fermi surface changes. The blue regions correspond to the weak
coupling gapped $p+ip$ phases, which are topological. Gray and maroon
regions: strong coupling phases. In mean field, the gapped $p+ip$
state wins over the non-chiral $p$ state in the strong coupling sector.  } }
\end{figure}

\begin{figure*}[t]
\label{fig4} \includegraphics[width=0.97\linewidth]{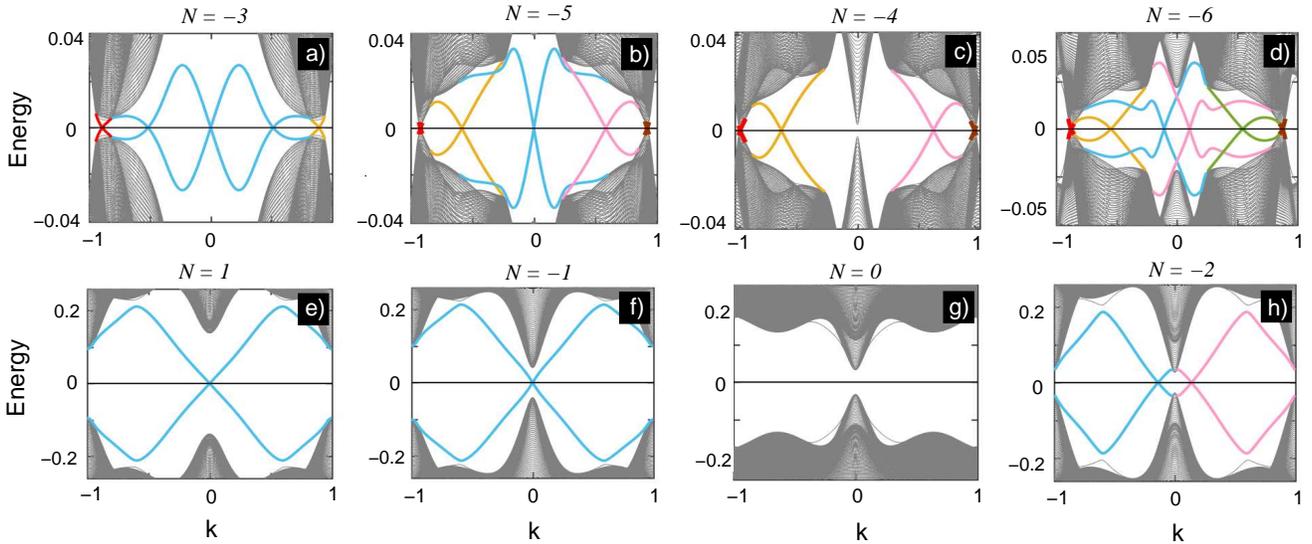}\caption{{\small{}{}(Color online) Majorana edge modes in the different topological
phases in the gapped $p+ip$ state. Energy units in eV. a) BdG Chern
number $\mathcal{N}=-3$ state, at $\mu=0.44$eV. b) $\mathcal{N}=-5$
at $\mu=0.33$eV; c) $\mathcal{N}=-4$ at $\mu=-0.49$ eV and d) $\mathcal{N}=-6$
at $\mu=-0.38$ eV in the weak coupling regime. The lower panels give
the corresponding phases in the strong coupling sector: e) $\mathcal{N}=1$
at $\mu=0.44$eV, f) $\mathcal{N}=-1$ at $\mu=0.33$eV; g) $\mathcal{N}=0$
at $\mu=-0.49$ eV,}\textbf{\small{} }{\small{}which is topologically
trivial and h) $\mathcal{N}=-2$ at $\mu=-0.38$eV. At the crossing
from the weak to strong coupling phases, when $g=g_{1c}(\mu)$, all
Chern numbers increase by 4.}}
\end{figure*}

When $\mu$ is in the immediate vicinity of the van Hove singularities,
$\bar{g}_{1c}$ abruptly drops towards zero. This singular behavior
suggests a crossover to exponential scaling when the Fermi surface
is nested at the van Hove singularities \cite{note2}. In that regime,
the phase transition is not quantum critical.\textcolor{black}{{} A
similar behavior is also observed in the $p$SC I phase near the van
Hove. At the saddle points, the order parameter can be broken into
four patches related by the $C_{4}$ symmetry of the bands combined
with odd angular momentum of the Cooper pairs. By symmetry, the free
energy written in a basis of $p_{x}$ and $p_{y}$ pairing states
is 
\begin{align}
\mathcal{F}(\Delta_{x}^{p},\Delta_{y}^{p}) & =\alpha(|\Delta_{x}^{p}|^{2}+|\Delta_{y}^{p}|^{2})+\beta\left(|\Delta_{x}^{p}|^{2}+|\Delta_{y}^{p}|^{2}\right)^{2}\nonumber \\
 & \qquad-\frac{\beta}{2}\left|(\Delta_{x}^{p})^{2}-(\Delta_{y}^{p})^{2}\right|^{2}+\mathcal{O}(\Delta^{6}),\label{eq:GL}
\end{align}
where $\alpha<0$ in the ordered state and $\beta>0$ \cite{note5}.
The last term favors coexistence between $\Delta_{x}^{p}$ and $\Delta_{y}^{p}$
with $\pm\pi/2$ phase difference ($p+ip$ phase), while the second
one favors a non-chiral $p$ state. The chiral and non-chiral phases
are exactly degenerate at the mean field level up to quartic order
terms in the expansion. Their degeneracy will likely be lifted by
fluctuations, which will be addressed elsewhere.}

In general, all the gapped chiral phases prevail over the gapless
one $(\Delta_{2}^{C}$). At small doping, the two critical couplings
of the weak and strong coupling phases merge ($\bar{g}_{1c}=g_{1c}$) below $|\mu|\lesssim0.6t$ and the
gapped phase has a first order phase transition to the normal state
at $g<g_{1c}(\mu)$ (see Fig 3b).

\emph{Topological phase transitions.} In 2D, spinless superconductors
with a bulk gap that breaks time reversal symmetry belong to the C
class in the ten-fold classification table \cite{class1,class2}.
The topological number in this class is defined by the BdG Chern number
\emph{$\mathcal{N}$, }which corresponds to the number of chiral Majorana
modes propagating along the edge \cite{Read,Thouless}.

In Fig. 3b, we explicitly calculate the Chern number 
\begin{equation}
\mathcal{N}=(i/2\pi)\int_{BZ}\text{d}^{2}\mathbf{q}\,\Omega_{z}(\mathbf{q})\label{eq:N}
\end{equation}
in the gapped state as a function of $\mu$ and intra-orbital coupling
$g_{1}$, with $\boldsymbol{\Omega}(\mathbf{q})=\nabla_{\mathbf{q}}\times\langle\psi_{n,\mathbf{q}}|\nabla_{\mathbf{q}}|\psi_{n,\mathbf{q}}\rangle$
the Berry curvature from the eigenstates of the BdG Hamiltonian at
the Fermi level, $|\psi_{n,\mathbf{q}}\rangle$. By changing the chemical
potential, the system shows a sequence of topological phase transitions.

In the weak coupling phase, shown in the blue areas in Fig. 3b, there
are up to five transitions separating different topological phases
with $\mathcal{N}=-4,\,-5,\,-6,\,-4,\,-5,$ and $-3$, in the range
of $-2t\leq\mu\leq2t=0.6$eV. The critical values of the chemical
potential where the system has a topological phase transition are
close to the energy of the van Hove singularities of the band (see
Fig. 1c) and coincide with the energies where the topology of the
Fermi surface changes. At those critical values, the superconducting
gap closes and the Chern number jumps by an integer number. The line
$g_{1}=\bar{g}_{1c}(\mu)$ separates the blue areas from the normal
region through continuous phase transitions. As anticipated, when
$|\mu|\lesssim0.6t=0.18$eV, $\bar{g}_{1c}=g_{1c}$, and the weak
coupling phases are suppressed. The singular behavior of $\bar{g}_{1c}(\mu)$
when $\mu$ is at the van Hove is not captured by the numerics shown
in Fig. 3 due to the smallness of the gap.

The solid curve separating the blue regions in Fig. 3b from the strong
coupling phases sets $g_{1c}(\mu)$, which describes a line of first
order phase transitions between different gapped phases. At this line,
the order parameter is discontinuous \cite{note3}, indicating the
onset of a topological phase transition as a function of $g_{1}$
for fixed $\mu$. In all cases, the Chern number changes accross the
$g_{1c}(\mu)$ line by $\Delta\mathcal{N}=4$. Deep in the strong
coupling regime (gray and maroon regions), for fixed $g>g_{1c}(\mu=0)$,
there are six topological phase transitions separating the phases
$\mathcal{N}=0,\,-1,\,-2,\,0,\,-1,\,1,\,0$ as a function of the chemical
potential. \textcolor{black}{The $\mathcal{N}=0$ phases (CSC) are
chiral but topologically trivial.} At the wide doping window $1.27t\lesssim\mu\lesssim2t=0.6$
eV, the elemental chiral topological superconducting phase with $\mathcal{N}=\pm1$,
and hence a single Majorana mode, can emerge at strong coupling.

\emph{Chiral Majorana edge states.} To explicitly verify the Chern
numbers for the different phases, we calculate the edge modes of the
gapped state in a two dimensional strip geometry with edges oriented
along the $(1,0)$ direction.

The plots in Fig. 4a$-$d (top row) show the evolution of the edge
modes in the weak coupling regime ($\bar{g}_{1c}<g<g_{c}(\mu)$) for
different values of $\mu$. The $\mathcal{N}=-3$ state shown in Fig.
4a has five edge modes in total, but only three modes that are topologically
protected, as indicted by the three different colors. The three modes
indicated in blue can be adiabatically deformed into a single zero
energy crossing at $k=0$, and hence count as a single topologically
protected mode. By decreasing the chemical potential into the contiguous
$\mathcal{N}=-5$ state (Fig. 4b), two of those modes become topologically
protected, raising the number of Majorana modes to five. By reducing
$\mu$ further into the $\mathcal{N}=-4$ state, the topology of the
Fermi surface changes drastically, forming gapped pockets of charge
around four Dirac nodes, indicated in Fig. 1c. Panel d shows the edge
modes of the $\mathcal{N}=-6$ state, for $\mu\lesssim-t=-0.3$eV.
The corresponding edge modes in the strong coupling regime ($g>g_{1c}(\mu)$)
with $\mathcal{N}=1,\,-1,\,0,$ and $-2$ are shown in the bottom
row of Fig. 4 (e$-$h). 

\emph{Pairing Mechanism.} Although it is difficult to reliably predict
a mechanism of superconductivity, at large doping and in the vicinity
of the van Hove singularities, where the DOS is very large, both phonons
and electronic interactions could be suitable candidates for a pairing
mechanism. We will not discuss the phonon mechanism, since it is conventional.

Electronic mechanisms typically provide attraction when the charge
susceptibility at the Fermi surface nesting vector $\mathbf{Q}$ satisfies
$\chi(\mathbf{Q})>\chi(0)$ \cite{Kohn}. When the chemical potential
$\mu$ is close to a Van Hove singularity, the electronic bands have
energy spectrum $\epsilon(\mathbf{q})=-\alpha q_{x}^{2}+\beta q_{y}^{2}$,
($0<\alpha\leq\beta)$ where $\mathbf{q}$ is the momentum away from
the saddle point. The susceptibility in the vicinity of the singularity
is logarithmic divergent, \emph{$\chi(0)=\frac{1}{2\pi^{2}}/\sqrt{\alpha\beta}\ln\left(\Lambda/\delta\mu\right)$
}with $\delta\mu$ the deviation away from the van Hove and $\Lambda\sim t$
an ultraviolet cut-off around the saddle point \cite{Gonzalez}. At
the nesting wavevector $\epsilon(\mathbf{q}+\mathbf{Q})=-\alpha p_{y}^{2}+\beta p_{x}^{2}$
, the susceptibility is 
\begin{equation}
\chi(\mathbf{Q})=c/(\alpha+\beta)\ln\left(\Lambda/\delta\mu\right),\label{eq:chi}
\end{equation}
where the constant $c=\frac{1}{\pi^{2}}\ln(\sqrt{\frac{\alpha}{\beta-\alpha}}+\sqrt{\frac{\beta}{\beta-\alpha}})$
is logarithmically divergent at the nesting condition $\alpha=\beta$
\cite{Pattnaik}. For the particular lattice Hamiltonian parametrization
taken from Ref. \cite{Cai}, the fitting of the bands around the van
Hove at $\mu=0.312$ eV has $\alpha\approx1.2$ and $\beta\approx1.7$.
That gives the ratio $\chi(\mathbf{Q})/\chi(0)\sim1.20$, suggesting
that a purely electronic mechanism of superconductivity is possible
\cite{Gonzalez,Guinea,Note4}. The high doping regime could in principle
be reached with gating effects for CrO$_{2}$ encapsulated in an insulating
substrate \cite{Mayorov} that preserves the roto-inversion symmetry
of the lattice. 

\emph{Acknowledgements.} BU acknowledges K. Mullen for discussions.
XD, KS and BU acknowledge NSF CAREER Grant No. DMR-1352604 for partial
support.


\begin{thebibliography}{10}
\bibitem{Yu}Yu. S. Dedkov, M. Fonine, C. König, U. Rüdiger, and G.
Güntherodt, Appl. Phys. Lett. \textbf{80}, 4181 (2002).

\bibitem{Soulen}R. J. Soulen, J. M. Byers, M. S. Osofsky, B. Nadgorny,
T. Ambrose, S. F. Cheng, P. R. Broussard, C. T. Tanaka, J. Nowak,
J. S. Moodera, A. Barry, and J. M. D. Coey, Science \textbf{282},
85 (1998).

\bibitem{Pickett}W. E. Pickett, Phys. Rev. Lett. \textbf{77}, 3185
(1996).

\bibitem{Mackenzie}A. P. Mackenzie and Y. Maeno, Rev. Mod. Phys.
\textbf{75}, 657 (2003).

\bibitem{Kallin}C. Kallin and J. Berlinsky, Rep. Prog. Phys. \textbf{79},
54502 (2016).

\bibitem{Maeno}Y. Maeno, S. Kittaka, T. Nomura, S. Yonezawa, K. Ishida,
J. Phys. Soc. Jpn. \textbf{81,} 011009 (2012).

\bibitem{Read}N. Read and D. Green Phys. Rev. B \textbf{61}, 10267
(2000).

\bibitem{Qi2}X.-L. Qi, T. L. Hughes, S. Raghu, and S.-C. Zhang, Phys.
Rev. Lett. \textbf{102}, 187001 (2009)

\bibitem{Qi}X.-L. Qi, and S.-C. Zhang, Rev. Mod. Phys. \textbf{83,
}1057 (2011).

\bibitem{Alicea}J. Alicea, Rep. Prog. Phys. \textbf{75}, 076501 (2012).

\bibitem{Lee}P. A. Lee, Science \textbf{346}, 545-546 (2014).

\bibitem{Sato}M. Sato and Y. Ando, Rep. Prog. Phys.\textbf{ 80},
076501 (2017).

\bibitem{Jackiw}R. Jackiw and P. Rossi, Nucl. Phys. B 190, 681 (1981).

\bibitem{Xu}J. P. Xu \emph{et al.}, Phys. Rev. Lett. \textbf{114},
017001 (2015).

\bibitem{Chung2}S. B. Chung, H. Bluhm, and E.-A. Kim, Phys. Rev.
Lett. \textbf{99}, 197002 (2007).

\bibitem{Jang}J. Jang, D. G. Ferguson, V. Vakaryuk, R. Budakian,
S. B. Chung, P. M. Goldbart, Y. Maeno, Science \textbf{331}, 186 (2011).

\bibitem{Fu}L. Fu and C. L. Kane Phys. Rev. Lett. \textbf{100}, 096407
(2008).

\bibitem{Akhmerov}A. R. Akhmerov, Johan Nilsson, and C. W. J. Beenakker,
Phys. Rev. Lett. \textbf{102}, 216404 (2009).

\bibitem{fu2}L. Fu, C. L. Kane, Phys. Rev. Lett. \textbf{102}, 216403
(2009).

\bibitem{Lutchyn}R. M. Lutchyn, J. D. Sau, and S. Das Sarma, Phys.
Rev. Lett. \textbf{105}, 077001 (2010).

\bibitem{Chung}S. B. Chung, H.-J. Zhang, X.-L. Qi, and S.-C. Zhang,
Phys. Rev. B \textbf{84}, 060510(R) (2011).

\bibitem{Li}J. Li, T. Neupert, Z. J. Wang, A. H. MacDonald, A. Yazdani,
B. A. Bernevig, Nat. Comm. \textbf{7}, 12297 (2016).

\bibitem{Qi3}X.-L. Qi, T. L. Hughes, and S.-C. Zhang, Phys. Rev.
B \textbf{82}, 184516 (2010).

\bibitem{He}Q. L. He \emph{et al.,} Science \textbf{357}, 294 (2017).

\bibitem{Schwartz}K. Schwarz, Journal of Physics F-Metal Physics
\textbf{16}, L211 (1986).

\bibitem{Katsnelson}M.I. Katsnelson, V.Yu. Irkhin, L. Chioncel, A.I.
Lichtenstein, R.A. de Groot, Rev. Mod. Phys. \textbf{80}, 315 (2008).

\bibitem{note1}While $t_{4}=0$ by the symmetry of the crystal, spin-orbit
coupling effects lead to a finite imaginary $t_{4}=it/8\approx\pm(i)0.036$
eV. See Ref. \cite{Cai}. This term opens a small gap of $\sim4$
meV at the Dirac nodes.

\bibitem{Cai}T. Cai, X. Li, F. Wang, S. Ju, J. Feng, and C.-D. Gong,
Nano Lett. \textbf{15}, 6434 (2015).

\bibitem{Furukawa} N. Furukawa, T. M. Rice, and M. Salmhofer, Phys.
Rev. Lett. \textbf{81}, 3195 (1998).

\bibitem{Nandkishore}R. Nandkishore, L. S. Levitov and A. V. Chubukov,
Nat. Phys. \textbf{8}, 158 (2012).

\bibitem{Kiessel} M. L. Kiesel, C. Platt, W. Hanke, D. A. Abanin,
and R. Thomale, Phys. Rev. B \textbf{86}, 020507(R) (2012).

\bibitem{note6}\textcolor{black}{The $p_{y}$ state is also allowed
by symmetry. All non-chiral combinations of $p_{x}$ and $p_{y}$
pairing symmetry have higher energy and are subdominant.}\textcolor{blue}{{} }

\bibitem{note2} For a more detailed analysis of the quantum critical
scaling, see supplementary materials.

\bibitem{note5}For a symmetry analysis, see supplementary materials. 

\bibitem{note3} The spectral gap does not close along the line of
first order topological phase transitions. The gap closes, however,
if $\Delta_{1}$ is virtually changed as a continuous parameter connecting
two topologically distinct ground states.

\bibitem{class1}A. P. Schnyder, S. Ryu, A. Furusaki, and A. W. W.
Ludwig, Phys. Rev. B \textbf{78}, (2008).

\bibitem{class2}S. Ryu, A. Schnyder, A. Furusaki, and A. Ludwig,
New J. Phys. \textbf{12}, 65010 (2010).

\bibitem{Thouless}D. J. Thouless, M. Kohmoto, M. P. Nightingale,
M. de Nijs, Phys. Rev. Lett. \textbf{49}, 405 (1982).

\bibitem{Kohn}W. Kohn, J. M. Luttinger, Phys. Rev. Lett. \textbf{15},
524 (1965).

\bibitem{Gonzalez}J. Gonzalez, Phys. Rev. B \textbf{78}, 205431 (2008).

\bibitem{Pattnaik}P. C. Pattnaik, C. L. Kane, D. M. Newns, C. C.
Tsuei, Phys. Rev. B\textbf{ 45}, 5714 (1992).

\bibitem{Guinea}F. Guinea, B. Uchoa, Phys. Rev. B \textbf{86}, 134521
(2012).

\bibitem{Note4} \textcolor{black}{This mechanism does not in principle
discard the possibility of particle-hole instabilities. In the square
lattice, they are known to be subleading to superconductivity in the
singlet channel. See Ref. \cite{Furukawa}.}

\bibitem{Mayorov}Mayorov \emph{et al.}, Nano Lett. \textbf{11}, 2396
(2011). 
\end{thebibliography}
\end{document}